\documentclass{vldb}
\pdfoutput=1
\usepackage{mathptmx}
\usepackage{inconsolata}
\usepackage{lineno}
\usepackage{graphicx}
\usepackage{xcolor}
\usepackage{listings}
\usepackage{balance}
\usepackage{booktabs}
\usepackage[justification=justified,singlelinecheck=false,font=small,format=plain,labelfont=bf,up,textfont=normal,up]{caption}
\usepackage{url}
\usepackage{hyperref}
\usepackage{cite}

\newif\iftechreport
\techreporttrue

\newcommand{\ignorethis}[1]{}

\newcommand{\Reals      }     {{\textrm{I\kern-0.18em R}}}

\newcommand{\change     } [1] {\mbox{{\footnotesize $\Delta$} \kern-3pt}#1}

\definecolor{darkred}{rgb}{0.7,0.1,0.1}
\definecolor{darkgreen}{rgb}{0.1,0.5,0.1}
\definecolor{cyan}{rgb}{0.7,0.0,0.7}
\definecolor{dblue}{rgb}{0.2,0.2,0.8}
\definecolor{maroon}{rgb}{0.76,.13,.28}
\definecolor{burntorange}{rgb}{0.81,.33,0}
\definecolor{royalpurple}{rgb}{0.47,.31,0.66}

\ifdefined\ShowNotes
  \newcommand{\colornote}[3]{{\color{#1}\bf{#2 #3}\normalfont}}
\else
  \newcommand{\colornote}[3]{}
\fi

\ifdefined\ShowNotes
  \newcommand{\num}[1]{{\color{red}\bf{#1}\normalfont}}
\else
  \newcommand{\num}[1]{#1}
\fi

\newcommand{\rekall}{\textsc{Rekall}}

\newcommand{\appendixortechreport}{\num{\iftechreport{Appendix}\else{technical report}\fi}}

\newcommand{\commercial}{\textsc{Commercial}}
\newcommand{\interview}{\textsc{Interview}}
\newcommand{\shotquery}{\textsc{Shot Detect}}
\newcommand{\shotscale}{\textsc{Shot Scale}}
\newcommand{\conversation}{\textsc{Conversation}}
\newcommand{\filmidiom}{\textsc{Film Idiom}}

\newcommand{\av}{\textsc{AV}}
\newcommand{\parking}{\textsc{Parking}}
\newcommand{\debugging}{\textsc{Debugging}}

\newcommand{\lam}{\mathsf{\lambda}}

\definecolor{mygray}{gray}{0.6}
\definecolor{keywordcolor}{rgb}{0.2,0.2,0.6}
\definecolor{keywordcolor2}{rgb}{0.15,0.46,0.1}
\definecolor{typecolor}{rgb}{0.17,0.56,0.68}
\definecolor{commentcolor}{gray}{0.3}
\definecolor{ratecolor}{rgb}{0.5,0.1,0.1}
\definecolor{stringcolor}{gray}{0.3}

\lstdefinelanguage{pseudo}{%
	morekeywords=[0]{Label,Interval,Metadata,rekall,join,map,filter,filter_against,union,group_by,minus,coalesce,matches,length},
    morekeywords=[1]{time_intersection,time_span,get_intersection_in_time,get_span_in_time,intersection_in_time,span_in_time,span},
	morekeywords=[2]{before,after,time_gap,start_time_equal,end_time_equal,time_equal,time_overlaps,time_before,time_after,overlaps,overlaps_in_time,height_greater,face_looking,lips_close,before_in_time,after_in_time,starts_inv_pred,equal_pred,finishes_inv_pred,meets,meets_inv,y_between,y_coords_close},
    morekeywords=[3]{or,and,def,for,in,any,len},
	morekeywords=[4]{interval,set,bool,string,list,label},
	morecomment=[l]{\#},
  stringstyle=\color{stringcolor},
  showstringspaces=false,
  morestring=[b]',
  morestring=[b]"
}

\newcommand{\code}[1]{\text{\lstinline[basicstyle=\ttfamily,mathescape]|#1|}}

\newcommand\todosilent[1]{}

\vldbTitle{Rekall: Specifying Video Events using Compositions of Spatiotemporal Labels}
\vldbAuthors{Daniel Y. Fu, Will Crichton, James Hong, Xinwei Yao,
Haotian Zhang, Anh Truong, Avanika Narayan, Maneesh Agrawala, Christopher R\'e,
Kayvon Fatahalian}
\vldbDOI{https://doi.org/10.14778/xxxxxxx.xxxxxxx}
\vldbVolume{12}
\vldbNumber{xxx}
\vldbYear{2019}

\iftechreport
  \def\endabstract{\if@twocolumn\else\endquotation\fi}
  \toappear{}
  \makeatletter
  \def\@copyrightspace{\relax}
  \makeatother
\fi

\newcommand{\AvgLift}{\num{6.5}}
\newcommand{\MaxLift}{\num{26.1}}

\begin{document}

\title{Rekall: Specifying Video Events using Compositions of Spatiotemporal Labels}
\numberofauthors{1}

\author{
\alignauthor
Daniel Y. Fu, Will Crichton, James Hong, Xinwei Yao,
Haotian Zhang, Anh Truong, Avanika Narayan, Maneesh Agrawala, Christopher R\'e,
Kayvon Fatahalian\\
       \affaddr{Stanford University}\\
       \email{\{danfu, wcrichto, james.hong, xinwei.yao, haotianz,
       anhlt92, avanika, maneesh,
       chrismre, kayvonf\}@cs.stanford.edu}
}

\date{30 July 1999}

\maketitle

\begin{abstract}
Many real-world video analysis applications require the ability to identify
domain-specific events in
video, such as interviews and commercials in TV news broadcasts, or action
sequences in film.
Unfortunately, pre-trained models to detect all the events of interest in video
may not exist, and training new models from scratch can be costly and
labor-intensive.
In this paper, we explore the utility of specifying new events in video
in a more traditional manner: by writing queries that compose outputs of
existing, pre-trained models.
To write these queries, we have developed \rekall, a library that exposes a
data model and programming model for compositional video event specification.
\rekall\ represents video annotations from different sources (object
detectors, transcripts, etc.) as spatiotemporal labels associated with
continuous volumes of spacetime in a video, and provides operators for
composing labels into queries that model new video events.
We demonstrate the use of \rekall\ in analyzing video from cable TV news
broadcasts, films, static-camera vehicular video streams, and commercial
autonomous vehicle logs.
In these efforts, domain experts were able to quickly (in a few hours to a day)
author queries that enabled the accurate detection of new events (on par with,
and in some cases much more accurate than, learned approaches) and to rapidly
retrieve video clips
for human-in-the-loop tasks such as video content curation and training data
curation.
Finally, in a user study, novice users of \rekall\ were able to
author queries to retrieve new events in video given just one hour of query
development time.
\end{abstract}

\section{Introduction}
\label{sec:introduction}

Modern machine learning techniques can
robustly annotate large video collections with basic information about
their audiovisual contents (e.g., face bounding boxes, people/object locations,
time-aligned transcripts).
However, many real-world video applications
require exploring a more diverse set of events in video.
For example, our recent efforts to analyze cable TV news
broadcasts required models to detect interview segments and commercials.
A film production team may wish to quickly find common segments such
as action sequences to put into a movie trailer.
An autonomous vehicle development team may wish to mine video
collections for events like traffic light changes or obstructed left turns
to debug the car's prediction and control systems.
A machine learning engineer developing a new model for video analysis
may search for particular scenarios to bootstrap model development or focus
labeler effort.

\begin{figure*}[ht!]
  \centering
  \includegraphics[width=7in]{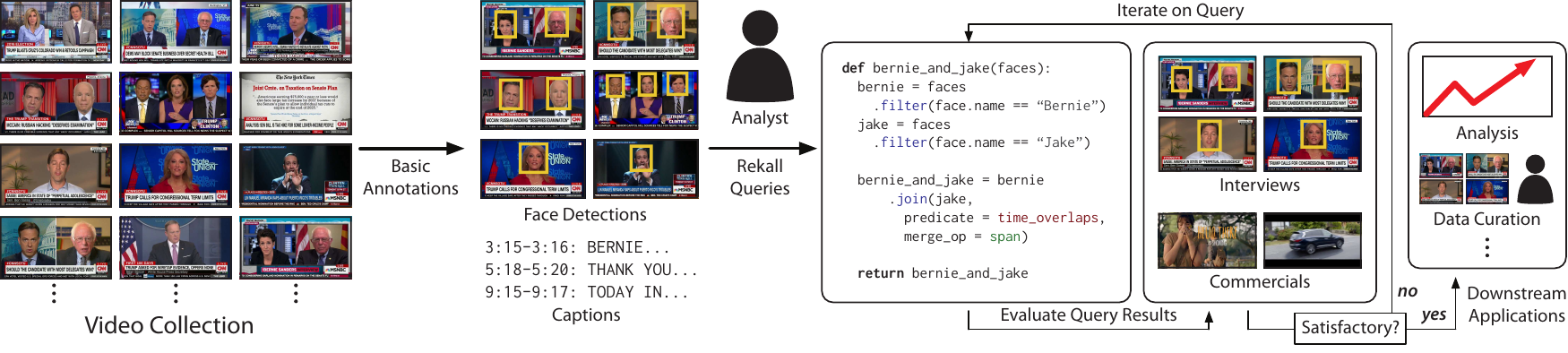}
  \caption{
  Overview of a compositional video event specification workflow.
  An analyst pre-processes a video collection to extract basic annotations about its 
  contents (e.g., face detections from an off-the-shelf deep neural
  network and audio-aligned transcripts).
  The analyst then writes and iteratively refines \rekall\ queries that compose
  these annotations to specify new events of interest, until query outputs
  are satisfactory for use by downstream analysis applications.
  }
  \label{fig:overview}
\end{figure*}

Unfortunately, pre-trained models to detect these domain-specific events often
do not exist,
given the large number and diversity of potential events of interest.
Training models for new events can be difficult and expensive,
due to the large cost of labeling a training set from scratch, and the
computation time and human skill required to then train an accurate model.
We seek to enable more agile video analysis workflows where
an analyst, faced with a video dataset and an idea for a new event of interest
(but only a small number of labeled examples, if any),
can quickly author an initial model for the event,
immediately inspect the model's results,
and then iteratively refine the model to
meet the accuracy needs of the end task.

To enable these agile, human-in-the-loop video analysis workflows,
we
propose taking
a more traditional approach: \textit{specifying
novel events in video as queries that programmatically
compose the outputs of existing, pre-trained models}.
Since heuristic composition does not require additional model training
and is cheap to evaluate, analysts can immediately inspect query
results as they iteratively refine queries to overcome 
challenges such as modeling complex event structure and dealing with
imperfect source video annotations (missed object detections,
misaligned transcripts, etc.).

To explore the utility of a query-based approach for detecting novel events
of interest in video,
we introduce \rekall, a library that exposes a data model and programming model for
\textit{compositional video event specification}.
\rekall\ adapts ideas from multimedia databases\,\cite{Donderler:2004:bilvideo,
Adali:1996:avis,Koprulu:2002:avis-st,hibino:1995:mmvis-tvql,oomoto:1993:ovid,
Kuo:1996:CVQL,Li1997ModelingOM} to the modern video analysis landscape, where
using the outputs of modern machine learning techniques allows for more
powerful and expressive queries,
and adapts ideas from complex event processing systems for
temporal data streams\,\cite{chandramouli2015trill,ApacheFlink,SiddhiQL} to the
spatiotemporal domain of video.

The primary technical challenge in building \rekall\ was defining the
appropriate abstractions and compositional primitives for users to write
queries over video.
In order to compose video annotations from multiple data sources that may be
sampled at different temporal resolutions
(e.g., a car detection on a single frame from a deep neural network,
the duration of a word over half a second in a transcript),
\rekall's data model adopts a unified representation of multi-modal video annotations,
the \textit{spatiotemporal label},  
that is associated with a continuous volume of spacetime in a video.
\rekall's programming model uses hierarchical composition of these labels to express 
complex event structure and define increasingly higher-level video events.

We demonstrate the effectiveness of compositional video event specification  
by implementing \rekall\ queries for
a range of video analysis tasks drawn from four application domains:
media bias studies of cable TV news broadcasts,
cinematography studies of Hollywood films, analysis of
static-camera vehicular video streams,
and data mining autonomous vehicle logs.
In these efforts, 
\rekall\ queries developed by domain experts with little prior 
\rekall\ experience achieved
accuracies on par with,
and sometimes significantly better than,
those of learning-based approaches (\AvgLift\ F1 points more accurate on
average, and up to \MaxLift\ F1 points more accurate for one task).
\rekall\ queries also served as a key video data retrieval
component of human-in-the-loop exploratory video analysis tasks.

Since our goal is to enable analysts to quickly retrieve novel events in video,
we also evaluate how well users are able to formulate \rekall\ queries for new
events in a user study.
We taught participants how to use \rekall\ with a one-hour tutorial, and then
gave them one hour to write a \rekall\ query to detect empty parking spaces
given the outputs of an off-the-shelf object detector.
Users with sufficient programming experience were able to write \rekall\
queries to express complex spatiotemporal event structures; these queries,
after some manual tuning (changing a single line of code) by an expert \rekall\
user to account for failures in the object detector, achieved near-perfect
accuracies (average precision scores above \num{94}).

To summarize, in this paper we make the following contributions:
\begin{itemize}
\item
We propose compositional video event specification as a
hu-man-in-the-loop
approach to rapidly detecting novel events of interest in video.

\item
We introduce \rekall, a library that exposes a data model and programming model
for compositional video event specification by adapting ideas from
multi-media databases and complex event processing over temporal data streams
to the modern video analysis landscape.

\item
We demonstrate the effectiveness of \rekall\ through analysis tasks across four
application domains, where domain experts were able to quickly author \rekall\
queries to accurately detect new events (on average \AvgLift\ F1 points more
accurate, and up
to \MaxLift\ F1 points more accurate, than learned approaches) and support
human-in-the-loop video
retrieval workflows.

\item
We evaluate how well novice users of \rekall\ are able to detect a novel event
in video given a one-hour tutorial and one hour of query development time
(average precision scores above \num{94} after tuning by an expert).
\end{itemize}
The rest of this paper is organized as follows:
Section~\ref{sec:overview} introduces an interview detection running example.
Section~\ref{sec:spatiotemporal} and~\ref{sec:compositions} use the running
example to introduce \rekall's data model and programming model.
Section~\ref{sec:tasks} introduces our application domains and analysis tasks,
and in Section~\ref{sec:evaluation} we evaluate the accuracy of the \rekall\
queries used to solve these tasks and evaluate the usability of \rekall\ for
video event specification.
Finally, we conclude with related work and discussion in
Sections~\ref{sec:related} and~\ref{sec:discussion}.
\iftechreport
Some supplemental videos can be found at \url{http://www.danfu.org/projects/rekall-tech-report/}.
\else
The technical report and supplemental videos can be found at
\url{http://www.danfu.org/projects/rekall-vldb2020/}.
\fi

\section{An Analysis Example}
\label{sec:overview}

To better understand the thought process underlying our video analysis
tasks, consider a situation where an analyst, seeking to understand sources of
bias in TV political coverage, wishes to tabulate the total time spent interviewing
a political candidate in a large collection of TV news video.
Performing this analysis requires identifying video segments that
contain interviews of the candidate.
Since extracting TV news interviews is a unique task,
we assume a pre-trained computer vision model is not available to the analyst.
However, it is reasonable
to expect an analyst does have access to widely available tools for detecting and
identifying faces in the video, and to the video's time-aligned text transcripts.

Common knowledge of TV news broadcasts suggests that interview segments tend to
feature shots containing faces of the candidate and the show's host framed
together, interleaved with headshots of just the candidate.
Therefore, a first try at an interview detector query might attempt to find
segments featuring this temporal pattern of face detections.
Refinements to this initial query might permit the desired pattern to
contain brief periods where neither individual is on screen (e.g., display of
B-roll footage for the candidate to comment on), or require parts of
the sequence to align with utterences of the candidate's name in the transcript
or common phrases like ``welcome" and ``thank for you being here."
As illustrated in Figure~\ref{fig:overview}, arriving at an accurate query for a dataset
often requires multiple iterations of the analyst reviewing query results and
adding additional heuristics as necessary until a desired level of accuracy is
achieved.

\begin{figure*}[ht!]
  \centering
  \includegraphics[width=7in]{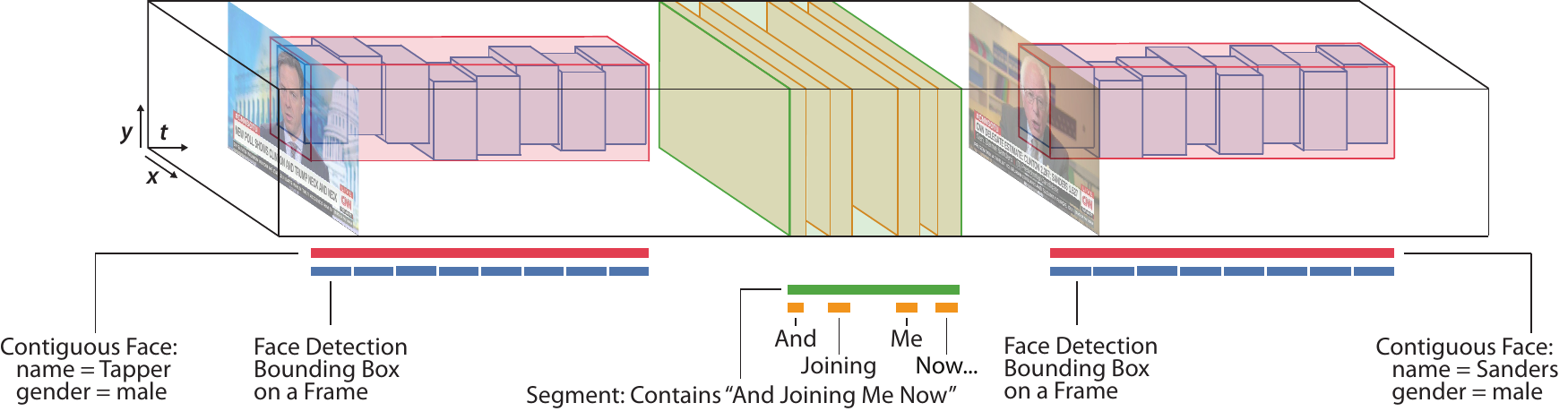}
  \caption{\rekall\ represents all video annotations, both basic annotations
  from computer vision models and annotations of more
  complex events, as labels associated with spatiotemporal intervals in
  the domain of a video. 
  \rekall's labels can be nested. 
  We illustrate two labels representing video segments where a face
  is continuously on screen (red) that contain labels corresponding 
  to per-frame face detections (blue), and one caption segment
  (green) that contains labels for individual words (orange).
  \label{fig:videovolume}}
\end{figure*}

Even in this simple example, a number of challenges emerge.
Annotations used as query inputs may be of different modalities and sampled at
different temporal rates (e.g., face detections are computed per frame,
transcript text is sub-second aligned).
Queries must be robust to noise in source annotations (e.g., missed face
detections, partially misaligned transcript data).
Lastly, to be sufficiently expressive to describe a range of events, the system must provide a
rich set of composition operators to describe temporal and (although not required in this example)
spatial relationships between annotations. 

The following sections describe \rekall's data model -- its representation of
multi-modal video annotation inputs -- and its programming model, the
operations available to queries for defining new video events in terms of these
inputs.

\iftechreport
\newpage
\fi

\section{Spatiotemporal Labels}
\label{sec:spatiotemporal}

To facilitate queries that combine information from a video sampled at different rates
and originating from different source modalities, \rekall\ adopts a unified
representation for all data: a \emph{spatiotemporal label} (or \emph{label}).  
Similar to how temporal databases associate records with an interval of time
designating their insertion to and deletion from the database\,\cite{jensen1994temporal},
each \rekall\ label is associated with a continuous, axis-aligned interval
of spacetime that locates the label in a video (a label is the equivalent of a
database record in \rekall's data model). 
For example, a face detected in a frame two minutes and ten seconds
into a 30~fps video with bounding box co-ordinates \{$x_1$: 0.2, $x_2$: 0.4, $y_1$: 0.2, $y_2$: 0.8\} 
(relative to the size of the frame) yields a label whose interval spans this
box in space and the range \{$t_1$: 2:10.0, $t_2$: 2:10.333\} in time.  
\rekall\ labels also optionally include metadata.
For example, a face detection label might include the name of the detected
individual:

\begin{lstlisting}[numbers=none]
face = Label(
  Interval=(t1: 2:10.0, t2: 2:10.333,
            x1: 0.2, x2: 0.4,
            y1: 0.2, y2: 0.8),
  Metadata={ identity: Bernie Sanders }
)
\end{lstlisting}

Figure~\ref{fig:videovolume} illustrates examples of 
labels generated for the TV news interview task introduced in
Section~\ref{fig:videovolume}.
Face detection performed on each frame yields labels (one per detected face,
blue boxes) that
span one frame of time (with the name of the individual as metadata).
The results of time-aligning the video's transcript yields a label per word
that extends for the length of the utterence (with the word as metadata, yellow
boxes).
Although words in a transcript are inherently temporal (and not spatial), they
can be lifted to the full spatiotemporal domain by assigning them the entire
space of the frame.
Most examples in this paper use a 3D (X,Y,T) video domain, although \rekall\
also supports intervals with additional spatial dimensions (e.g., labels in a
LIDAR point cloud video exist in a 4D domain).

To echo the hierarchical nature of information derived from a video, labels in
\rekall\ queries can be hierarchical (a label's metadata can be a list of labels.)
For example, a set of temporally continuous face detections of the same
individual at the same location on screen might be grouped into a single label
representing a segment where the individual is on screen.
The \num{red} boxes in Figure~\ref{fig:videovolume} indicate segments
where anchor Jack Tapper or guest Bernie Sanders are on screen.
The figure also shows a label corresponding to the phrase ``And joining me now"
that contains labels for the constituent words (green box).
Many of the queries described in the subsequent sections construct multi-level
hierarchies of labels.
For example, in TV and film videos, frames can be organized into shots, and
shots can be grouped into scenes (or news segments).

\section{Composing Labels}
\label{sec:compositions}

\begin{figure*}[ht!]
  \centering
  \includegraphics[width=7in]{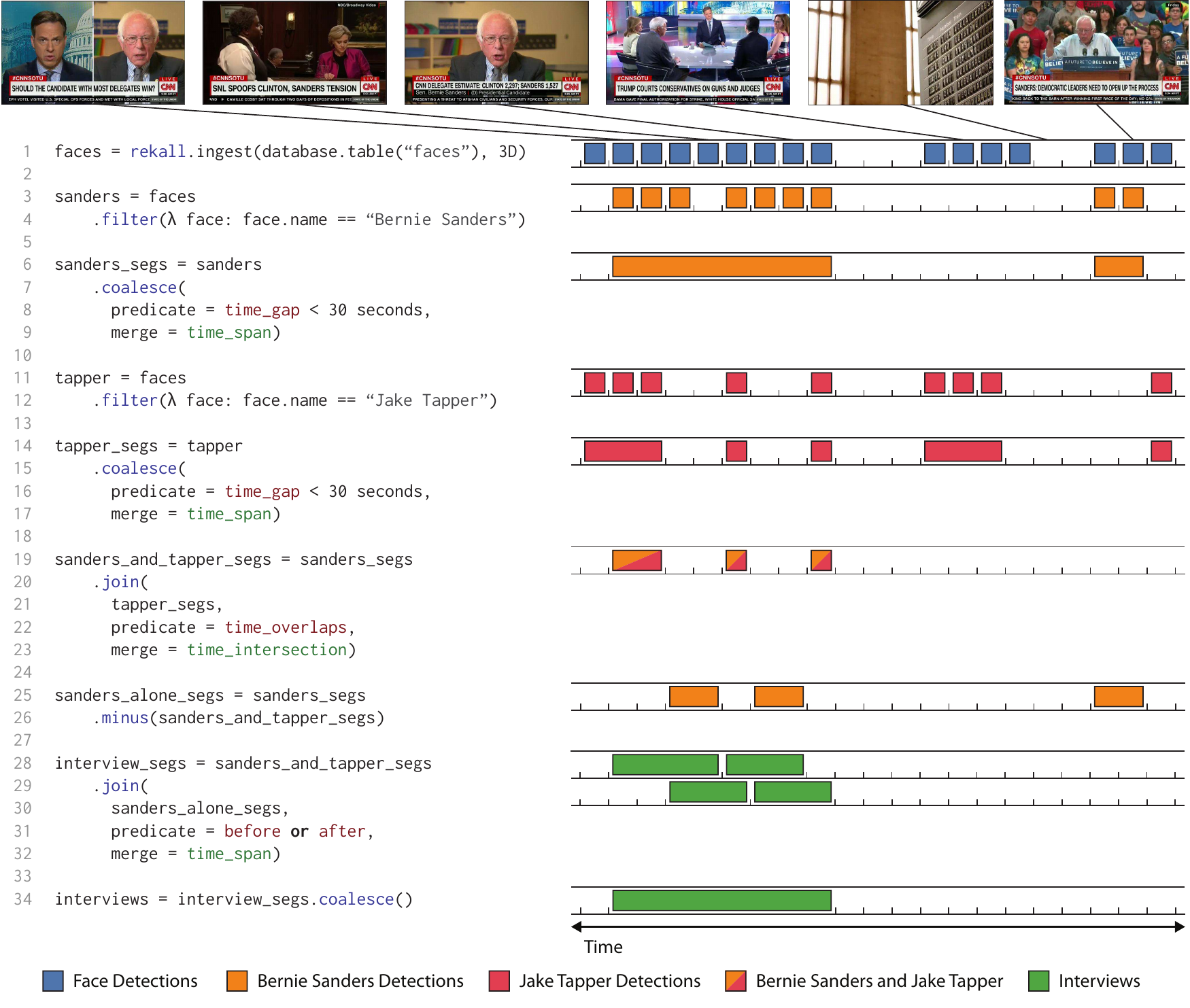}
  \caption{
  Left: A \rekall\ query for detecting interviews of Bernie Sanders by Jack Tapper.
  Right: A visual depiction of the intermediate label sets 
  produced over the course of execution.
  Blue: face detections; Orange: Bernie Sanders; Red: Jake Tapper; Orange and
  Red: Bernie Sanders and Jake Tapper; Green: interview segments.
  }
  \label{fig:compositions}
\end{figure*}

\rekall\ queries define how to compose existing labels into new labels
that correspond to instances of new events in a video.
In this section we describe the label composition primitives
available to \rekall\ queries.
To aid description, Figure~\ref{fig:compositions} provides code
for the TV news interview detection task from Section~\ref{sec:overview},
which will be used as the running example throughout this section.

\subsection{Label Sets}
\label{sec:eventsets}

\rekall\ uses sets of labels to represent all occurrences of an event in a
video (a label set is equivalent to a database relation).
A \rekall\ query consists of operations that produce and consume
sets of labels (all \rekall\ operations are closed on sets of labels).
For example, Line 1 of Figure~\ref{fig:compositions} constructs
an initial label set from a database table containing all face detections from
a video.
The variable \code{faces} is a set containing one label for each detected face.
The result of the query is the set \code{interviews}, which
contains one label corresponding to each interview segment in the video.

\rekall\ provides standard data-parallel operations \code{map},
\code{filter}, and \code{group_by} to manipulate label sets.
For example, lines \num{3} and \num{11} filter \code{faces} according to
the person's name (pre-computed using off-the-shelf identity
recognition tools\,\cite{Rekognition} and stored as label metadata) to produce
label sets containing only detections of Jake Tapper (\code{tapper}) and Bernie
Sanders (\code{sanders}).  

\begin{figure*}[ht!]
  \centering
  \includegraphics[width=7in]{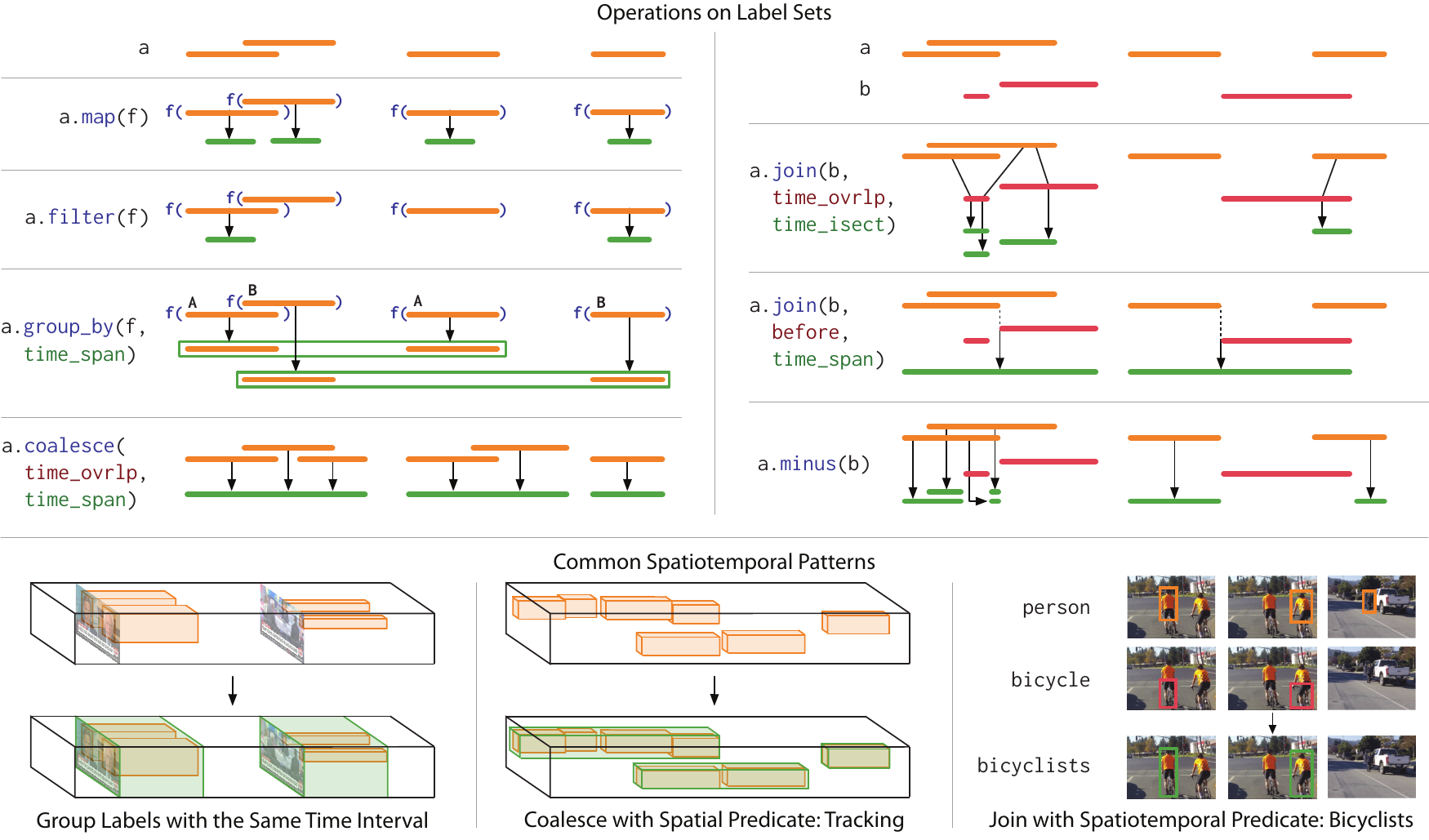}
  \caption{
  Semantics of \rekall\ operations on label sets.
  The functions shown are a sample of \rekall's complete library of operators,
  intended to provide visual intuition for how a few relevant operators work.
  Top half: label set operations, depicted on sets of one-dimensional temporal
  labels.
  Bottom half: depictions of common spatiotemporal operations on 
  multi-dimensional labels.
  }
  \label{fig:all_set_operations}
\end{figure*}

Figure~\ref{fig:all_set_operations} illustrates the behavior of various \rekall\
label set operations.
\code{map} can be used to manipulate the metadata or the spatiotemporal interval
associated with labels (e.g., shrink the interval as shown in the figure).
\code{group_by} is one mechanism by which \rekall\ queries construct nested labels.
For example, Figure~\ref{fig:all_set_operations}, bottom-left shows the use of
\code{group_by} to reorganize a set of four face detection labels into a set of
two labels that each contain a set of labels corresponding to faces in the same frame. 

\subsection{Coalesce: Recursive Label Merging}

Many video analysis tasks involve reasoning about sequences of labels or
about spatially adjacent labels.
For example, in the TV interviews query,
it is preferable to reason about continuous segments
of time where a person is on screen
(``a segment containing Bernie Sanders on screen, followed by one with Jake Tapper"),
rather than individual frame face detections.

\rekall's \code{coalesce} operator serves to merge an unbounded number of
fine-grained labels in close proximity in time or space into new labels that
correspond to higher-level concepts.
\code{coalesce} is parameterized by a query-specified \emph{label merge
predicate}, which determines whether a pair of labels should be merged, and a
\emph{label merge function} that specifies how to create a new label as a
result of a merge.
\code{coalesce} recursively merges labels in its input set (using the merge
function), until no further merges are possible (as determined by the merge
predicate).

Lines \num{6} and \num{14} in Figure~\ref{fig:compositions} demonstrate use of
\code{coalesce} to merge label sets of per-frame face detections into label sets
corresponding to sequences of time when Bernie Sanders and Jake Tapper are on screen.
In this example, the query uses a merge predicate that merges all input labels that
lie within 30~seconds of each other (not just temporally adjacent labels) and
the merge function combines the two labels to create a new label whose
spatiotemporal interval spans the union of the intervals of the two inputs.
As a result, the resulting labels, which correspond to video segments 
showing the interviewer or interviewee on screen, may contain brief cuts away
from the individual.

\code{coalesce} serves a similar role as ``repeat" operators in prior multimedia database
systems\,\cite{Donderler:2004:bilvideo,
Adali:1996:avis,Koprulu:2002:avis-st,hibino:1995:mmvis-tvql,oomoto:1993:ovid,
Kuo:1996:CVQL,Li1997ModelingOM}, or as a regex Kleene-star operator on point events
in event processing systems\,\cite{ApacheFlink,chandramouli2015trill}.
However, it is a general mechanism that provides queries the flexibility to build
increasingly higher levels of abstraction in custom ways.
For example, it is common to use \code{coalesce} with a spatiotemporal proximity
predicate to build labels that correspond to ``tracks" of an object
(Figure~\ref{fig:all_set_operations}, bottom-center) or to smooth over noise in
fine-grained labels (e.g., flicker in an object detector).
We document a variety of use cases of \code{coalesce} in Section~\ref{sec:tasks}.

\iftechreport
\newpage
\fi
\subsection{Joins}
\label{sec:joins}

Video analysis tasks often involve reasoning about multiple concurrent label
streams; \rekall\ provides standard join operators to combine multiple streams
together and construct new labels from spatiotemporal relationships between
existing labels.
\rekall's inner \code{join} is parameterized by a join predicate
and a label merge function that specifies how to merge matching pairs of labels
(like \code{coalesce}).
For example, line~\num{19} in Figure~\ref{fig:compositions} uses
\code{join} with a temporal overlap predicate (\code{time_overlaps}) and a
temporal intersection merge function (\code{time_intersection}) to generate a label set
(\code{sanders_and_tapper_segs}) corresponding to times when both Sanders and Tapper are on screen.
Line~\num{28} in the pseudocode uses a different join predicate (\code{before or after}) 
and merge function (\code{time_span}) to construct labels for segments where
Sanders is on screen directly before or directly after a segment containing both
Sanders and Tapper.

As shown at right in Figure~\ref{fig:all_set_operations},
\rekall\ provides a standard library of common spatial and temporal predicates
such as the Allen interval operations\,\cite{Allen:1983:interval} and their
2D spatial analogues\,\cite{cohn1997spatialcalculus}.
The bottom-right of the Figure illustrates the use of a join with a 
predicate (``above") to construct labels for bicyclists from label sets of
bicycle and person detections.

\rekall's \code{minus} operation is an anti-semi-join (similar to Trill's
\code{WhereNotExists}\,\cite{chandramouli2015trill}) that takes two label sets
and removes intervals associated with labels in the second set from the intervals
of labels in the first.
For example, Line~\num{10} in Figure~\ref{fig:compositions} uses \code{minus}
to construct labels for segments when Bernie Sanders is on screen alone.
Like \code{join}, \code{minus} is parameterized by a predicate that determines
which labels are matched by the operation.
The empty parking space detection query shown in Figure~\ref{lst:parking}
illustrates the use of \code{minus} with a spatiotemporal predicate that only
matches labels when their interval intersection to union ratio (IOU) exceeds a
manually-set threshold.

\section{Applications}
\label{sec:tasks}

\begin{table*}[t!]
    \centering
    \small
    \begin{tabular}{@{}llllcccc@{}}
        \toprule
        \textbf{Task}                               & \textbf{Application(s)}  & \textbf{Data Sources}                    & \textbf{Description}                           \\ \midrule 
        Commercial Detection (\commercial)          & TV News                  & Histograms, transcripts                  & Detect all commercial segments                 \\
        Interview Detection (\interview)            & TV News                  & Face detections                          & Detect all interviews with a particular guest  \\
                                                    &                          &                                          & (e.g., Bernie Sanders)                         \\
        Shot Transition Detection (\shotquery)      & Film                     & Histograms, face detections              & Detect every shot transition                   \\
        Shot Scale Classification (\shotscale)      & Film                     & Face detections, pose estimations        & Classify the scale of each shot                \\
        Conversation Detection (\conversation)      & Film                     & Face detections, face embeddings         & Detect all conversations                       \\
        Film Idiom Mining (\filmidiom)              & Film                     & Face detections, transcripts,            & Detect various film idioms -- reaction shots,  \\
                                                    &                          & histograms                               & action sequences, establishing shots, etc.     \\
        Empty Parking Space Detection (\parking)    & Static-Camera Feeds      & Object detections                        & Detect all empty parking spots                 \\
        AV Log Mining (\av)                         & AV                       & Object detections from cameras,          & Detect various rare events from autonomous    \\
                                                    &                          & LIDAR                                    & vehicle logs                                  \\
        Upstream Model Debugging (\debugging)       & TV News, Film, AV,       & Model outputs                            & Detect errors in model outputs                 \\
                                                    & Static-Camera Feeds      &                                          &                                                \\
        \bottomrule \\
        \vspace{-2em}
        \end{tabular}
        \caption{
        Nine representative tasks from \rekall\ deployments for analysis of a
        large collection of cable TV News, cinematographic studies of Hollywood
        films, analysis of static-camera vehicular video feeds, and data mining
        commercial autonomous vehicle logs.
        }
    \label{table:tasks}
\end{table*}

We have used \rekall\ to write queries needed for video analysis tasks
in several domains: media bias studies of TV news broadcasts,
analysis of cinematography trends in feature length films,
event detection in static-camera vehicular video streams,
and data mining the contents of autonomous vehicle logs.
In many cases, these queries have been used to automatically label large video
collections for statistical analysis;
in other cases, \rekall\ queries have also proven to be a valuable mechanism
for retrieving video clips of interest in scenarios
that involve human-in-the-loop analysis tasks.

The remainder of this section provides further detail on how \rekall's programming model
was used to author queries used in these application domains.
Table~\ref{table:tasks} enumerates several tasks from \rekall's deployments,
and includes the basic annotations used as input to \rekall\ queries that
perform these tasks.
Code listings for \shotscale\ and \parking\ are provided in this section, and
code listings for additional queries are provided in the \appendixortechreport.

\subsection{Analysis of Cable TV News Media}
We have used \rekall\ queries as
part of an ongoing study of representation and bias in over 200,000 hours of
U.S. cable TV news (CNN, MSNBC, FOX) between 2010 and 2018.
This effort seeks to analyze differences in screen time afforded to individuals
of different demographic groups, and asks questions such as  \emph{``Did
Donald Trump or Hillary Clinton get more interview screen time in the months
before the 2016 election?''} or \emph{``How much more screen time is given to
male presenting vs. female presenting hosts?"}
To answer these questions, screen time aggregations needed to be 
scoped to specific video contexts, such as interview segments, and needed to
exclude commercials. Thus, a key challenge involved developing queries for
accurately detecting commercial segments (\commercial) and segments featuring
interviews with specific political candidates (\interview).

A simplified version of the interview detection algorithm was described in
Section~\ref{sec:compositions};
in practice, we extend the algorithm to find interviews between \textit{any}
individual known to be a cable TV news host and a specified guest 
(we used Jake Tapper for expositional simplicity;
modifying the query in Figure~\ref{fig:compositions} to find candidate interviews
with any host is a one-line change).
In the TV news dataset, commercial segments often begin and end with short
sequences of black frames and often have mixed-case or missing transcripts
(news broadcasts typically feature upper-case caption text).
The \commercial\ query exploits these dataset-specific signals
to identify commercial
segments.
More details can be found in the \appendixortechreport.

\begin{figure*}[ht!]
  \centering
  \includegraphics[width=7in]{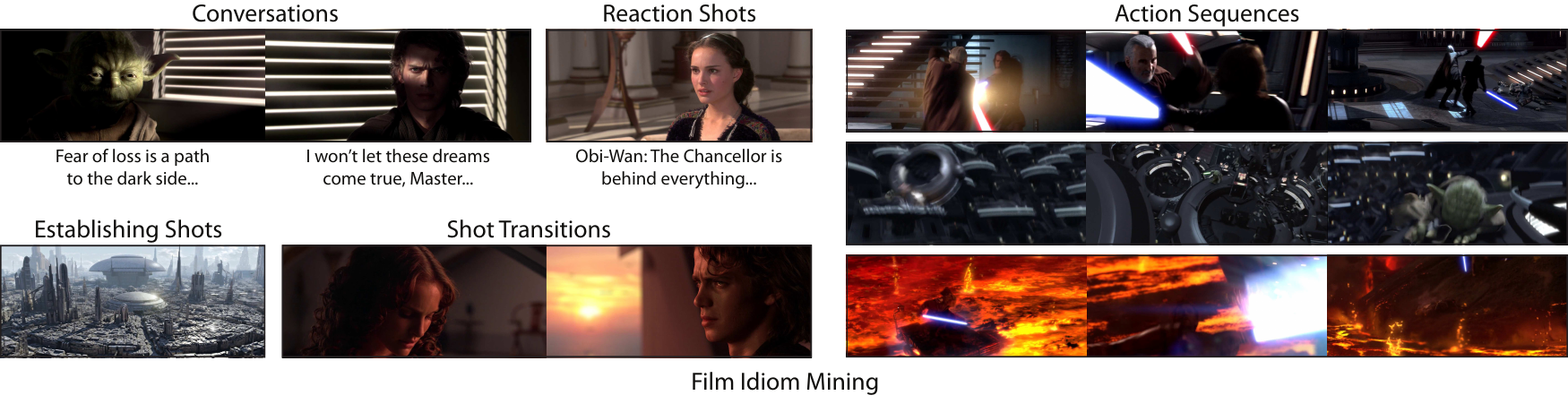}
  \caption{
  Examples of film idioms extracted from
  \textit{Star Wars: Episode III - Revenge of the Sith} using \rekall\ queries.
  }
  \label{fig:film_idioms}
\end{figure*}

\begin{figure}[h]
\centering
\includegraphics[width=3.33in]{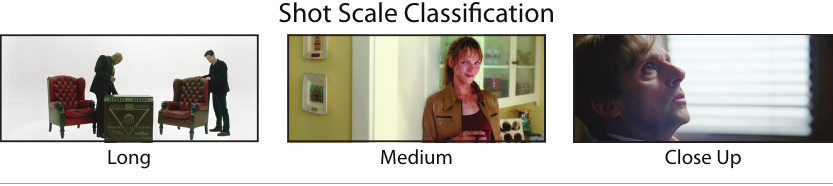}
\begin{lstlisting}
faces = rekall.ingest(database.table("faces"), 3D)
poses = rekall.ingest(database.table("poses"), 3D)
shots = rekall.ingest(database.table("shots"), 1D)

faces_per_frame = faces
  .group_by($\lam$ obj: (obj["t1"], obj["t2"]), span)
poses_per_frame = poses
  .group_by($\lam$ obj: (obj["t1"], obj["t2"]), span)

frame_scales_face = faces_per_frame
  .map(frame_scale_face)
frame_scales_pose = poses_per_frame
  .map(frame_scale_pose)

frame_scales = frame_scales_face
  .union(frame_scales_pose)
  .group_by($\lam$ frame: (frame["t1"], frame["t2"]), span)
  .map($\lam$ frame: take_largest(frame.nested))

shot_scales = frame_scales
  .join(
    shots,
    predicate = time_overlaps,
    merge = time_span)
  .group_by($\lam$ shot: (shot["t1"], shot["t2"]), span)
  .map($\lam$ shot: mode(shot.nested))
\end{lstlisting}
\caption{
A query for classifing the scale of a cinematic shot 
(as long, medium, or close up) based on the size of faces
and human pose estimates detected in frames.
The query uses nested labels to group per-detection
estimates of scale by frame, then pools per-frame
results by shot to arrive at a final estimate.}
\label{lst:shot_scale}
\end{figure}

\subsection{Film Cinematography Studies}
\label{sec:tasks_film}
We have analyzed a collection of $589$ feature-length films
spanning 1915-2016 to explore questions about
cinematographic techniques, and how their use has changed over time
(e.g., \emph{``How has the pace or framing of films changed over the past century?",
or ``How much screen time do conversations take in
films?"}\,\cite{Brunick:2013:lowlevel,
Cutting:2011:quickfastdark,Cutting:2016:pace,Cutting:2015:increasedpace,
Cutting:2016:facialexpression,Cutting:2015:framing}).
Representative queries are depicted in Figure~\ref{fig:film_idioms},
and include shot transition detection (\shotquery,
partitioning a video into segments of continuously recorded footage),
classifying shots by the relative size of the actors to the size of the frame
(\shotscale),
and conversation detection (\conversation).
Our film analyses also required queries for retrieving segments exhibiting
common film cinematography idioms such as action sequences or wide-angle scene
``establishing shots" for video clip content curation (\filmidiom).

As one example of a query used in this effort,
Figure~\ref{lst:shot_scale} provides code for \shotscale,
which classifies each shot
in a video as ``long", ``medium", or ``close up" based on the size
of actors on screen.  Inputs to this query include a label
set for all film shots (\code{shots}), as well as label sets for per-frame
face detections \code{faces} and actor body poses (\code{poses}).
Each shot contains multiple frames, each of which contains zero or more actors.
The challenge of this query is to use the relative sizes of each actor in a
frame to estimate the scale for the frame, and then use per-frame scale
estimates to estimate the scale of the shot.
The \rekall\ query echoes this nested structure using hierarchical labels to
estimate the scale of a shot.

The query first estimates the scale based on each face detection or pose
detection in a frame.
Lines~\num{10} and~\num{12}
estimate the
scale based on the relative size of face bounding boxes or pose skeletons,
using \code{frame_scale_face} or \code{frame_scale_pose}, respectively.
The query then aggregates these estimates into a single label set for each
frame, retaining the largest estimate (\code{take_largest}) from all detected
faces and poses (lines~\num{15-18}).
Finally, the query identifies the frames contained within each shot and
classifies the shot's scale as the mode of the scales computed for each
constituent frame (lines~\num{20-26}).

We provide details about the other cinematography queries in the \appendixortechreport.
These queries use rapid changes in video frame pixel color histograms and face
bounding box locations to detect shot boundaries,
identify patterns of shots where the same two individuals appear over an
extended period of time as a signal for conversations,
and use patterns in length or scale of consecutive shots to identify common film idioms.

\textbf{Human-in-the-loop Movie Data Exploration}.
In addition to conducting film analyses, we have also used \rekall\ queries as
a tool for exploring films and curating content needed for
video editing tasks like making video supercuts or movie
trailers\,\cite{smith2017harnessing,brachmann2007automatic}.
These video editing tasks require finding clips in a film that embody common
film idioms such as action shots, ``hero shots" of the main characters,
or wide-angle establishing shots. 

We authored \rekall\ queries for several different types of film idioms, 
and provided query results to a video editor who selected final clips to use in 
the movie trailer.
Query recall in this task was more important than precision, since the editor
could quickly result sets to select a small number of desirable clips.
We demonstrated this workflow on \textit{Star Wars: Episode III - Revenge of
the Sith}.
Once queries were written, the entire mining and trailer construction process
took less than four hours.
The full list of film idioms we mined for this effort can be found in the
\appendixortechreport.
The movie trailer, a selection of supercuts, and examples of other film idioms
can be found at \url{http://www.danfu.org/projects/rekall-tech-report/}.

\begin{figure}[h!]
\centering
\includegraphics[width=3.33in]{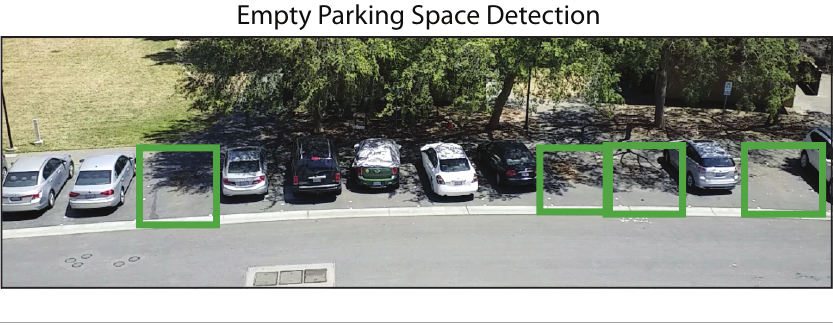}
\begin{lstlisting}
objects = rekall.ingest(database.table("objects"), 3D)

parking_spots = objects
  .filter($\lam$ obj: obj["t1"] == 0 and obj.class == "car")
  .map($\lam$ spot: spot with "t1" = 0 and "t2" = video_end)

vehicles = objects
  .filter($\lam$ obj: obj.class in ["car", "truck"])

empty_spot_candidates = parking_spots
  .minus(vehicles,
    predicate = $\lam$ spot, car: iou(spot, car) > 0.25)

empty_parking_spots = empty_spot_candidates
  .coalesce(
    predicate = $\lam$ spot, car: iou(spot, car) == 1,
    merge = span)
  .filter($\lam$ spot: spot["t2"] - spot["t1"] > 60 * 4)
\end{lstlisting}
\includegraphics[width=3.33in]{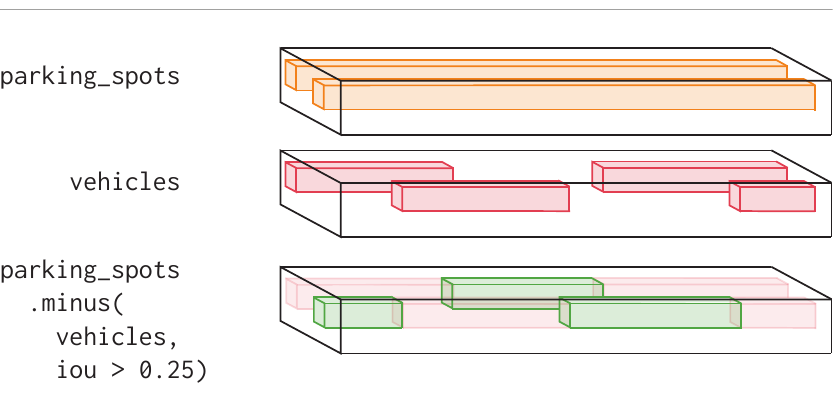}
\caption{
A query for detecting empty parking spaces using only
object detection results in a fixed-camera feed.
Potentially empty spots are obtained by subtracting the current
frame's car intervals (red) from a set of intervals corresponding
to all spots (orange).  To account for errors in an object detector,
a parking spot must be car free for a continuous period of time
(4~minutes) to be counted as a ``free spot".}
\label{lst:parking}
\end{figure}

\iftechreport
\else
\fi
\subsection{Static-Camera Vehicular Video Streams}
Inspired by recent vision applications in
the wild\,\cite{geitgeyparkingspot}, \parking\ detects the time
periods where parking spots are free in a fixed-camera video
feed of a parking lot.
The query, whose code is given in Figure~\ref{lst:parking}, uses only
labels produced by an off-the-shelf object detector run on
the video stream, and is based on two simple heuristics:
a parking spot is a spatial location where a car is stationary for a
long period of time, and an empty parking spot is a parking
spot without a car in it.

For simplicity, the query in Figure~\ref{lst:parking} assumes that all
parking spaces are taken at the start of the video
(this assumption can be relaxed by extending the query to identify
regions that are occupied by cars for a significant period of time at any point
in the video).
The query extends the car detections at the start of the video to the entire
video to construct a label set of parking spots (lines~\num{3-5}), and then
subtracts out car and truck detections (lines~\num{7-12}).
The predicate \code{iou} (intersection-over-union) provided to the \code{minus} operator ensures
that the operation only removes times when the parking spot is
completely filled (and not when vehicles partially overlap in pixel space when
passing by).
The behavior of the spatiotemporal \code{minus} operation, and the
surviving labels that correspond to empty parking spots
(green intervals), is illustrated at the bottom of Figure~\ref{lst:parking}.
Finally, to avoid errors due to missed object detections, the query uses
\code{coalesce} to construct consecutive empty spot detections, and removes
detections that exist for less than four minutes (lines~\num{14-18}).

\subsection{Mining Autonomous Vehicle Logs}
\rekall\ queries are used at a major autonomous vehicle company to mine for
rare, but potentially important, events in autonomous vehicle logs (\av).  
Queries are used to identify traffic light changes in quick succession,
as well as turns by other vehicles near the autonomous vehicle.
Images from the traffic light sequences are sent to human labelers for ground
truth annotation, 
which may reveal correct detector behavior (a sequence of green-to-yellow-to-red transitions),
major failures of the traffic light color detector (thus creating new labeled
supervision for an active learning loop),
or important rare situations to document such as a yellow flashing warning light.
Images from vehicle turn sequences, on the other hand, are sent to labelers to
label turn signals, and to validate the car's prediction and control systems.
As in the video editing case, since clips matching these queries are
subsequently passed on to human labelers for review, \rekall\ queries
serve as a filter that focuses human effort on examples that are most
likely to be important.

\begin{figure}[ht!]
  \centering
  \includegraphics[width=3.33in]{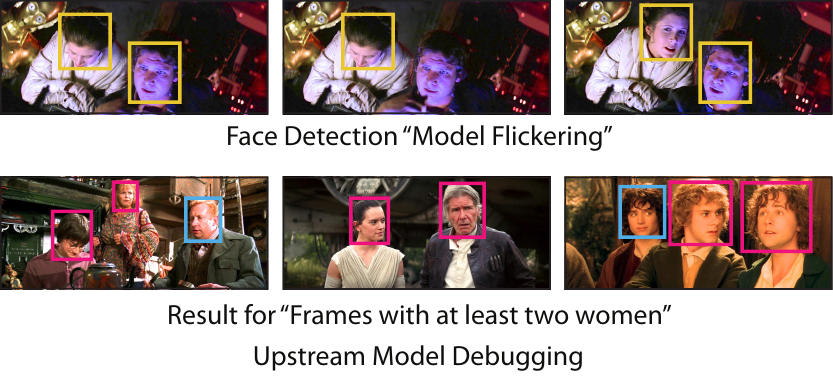}
  \caption{
  Two examples using \rekall\ queries to debug computer vision models during
  our film cinematography studies.
  Top: Face detections ``flickering" in and out of existence for a single frame
  at a time.
  Here, the face detector fails to detect Han Solo's face in the middle frame,
  even though his face appears in the same location and the same place in the
  surrounding two frames.
  Bottom: The result of a query for frames with two women based on the results
  of an off-the-shelf gender classifier.
  Gender balance is so poor in fantasy movies that most of these examples are
  false positives.
  Male classifications are shown in blue; female classifications are shown in
  red.
  Harry Potter, Han Solo, Merry and Pippin have all been misclassified.
  }
  \label{fig:model_debugging}
\end{figure}

\subsection{Model Debugging}
In all the projects using \rekall, queries have been used to identify errors in
the output of pre-trained models.
A common example is shown in Figure~\ref{fig:model_debugging}, where a
\rekall\ query is used to detect false negatives in a face detector
from a lack of temporal coherence in its output
(Han Solo's face is not detected in the middle frame).
Once users identify such errors, they often use \code{coalesce} to
smooth over the errors to improve end results (\parking\ smooths over errors in the
object detector, for example, increasing accuracy by \num{19.8} AP points over
a version of the query that does not).
The \code{coalesce} operations in the \interview\ algorithm and aggregation
steps in the \shotscale\ algorithm played similar roles in those queries.

During our film study efforts, a query for detecting frames with at least
two woman faces surfaced errors in the gender classifier, since scenes with
multiple women in Hollywood films are rare due to a bad gender imbalance.
Most scenes retrieved by the query were false positives due to incorrect
gender classification of faces.  
As a result, we subsequently replaced the face gender classifier
with a better model.
We note that our efforts searching for potentially anomalous patterns in trained model
output is similar to recent work on improving models using model
assertions\,\cite{kang2018model}.

\section{Evaluation}
\label{sec:evaluation}
\begin{table*}[ht]
    \centering
    \small
    \begin{tabular}{@{}rrrrrccc@{}}
        \toprule
        \multicolumn{1}{c}{} & \multicolumn{4}{c}{\textbf{Method}} \\ \cmidrule(l){2-5} 
        \textbf{Task}        & \textbf{ResNet-50 Image Classification} & \textbf{ResNet-50 Image Classification + Smoothing} & \textbf{Conv3D Action Recognition} & \textbf{\rekall} \\ \midrule 
        \interview           & 80.0 $\pm$ 3.4                & 87.3 $\pm$ 2.4                            & 17.7 $\pm$ 18.3             & \textbf{95.5}   \\
        \commercial          & 90.9 $\pm$ 1.0                & 90.0 $\pm$ 0.9                            & 88.6 $\pm$ 0.4              & \textbf{94.9}   \\
        \conversation        & 65.2 $\pm$ 3.5                & 66.1 $\pm$ 3.5                            & \textbf{79.4 $\pm$ 2.3}     & 71.8            \\
        \shotquery           & --                            & --                                        & 83.2 $\pm$ 1.0              & \textbf{84.1}   \\
        \shotscale           & 67.3 $\pm$ 1.0                & 68.1 $\pm$ 1.2                            & 70.1 $\pm$ 0.8              & \textbf{96.2}   \\ \midrule
        \textbf{}            & \textbf{Faster R-CNN Object Detection} & \textbf{} & \textbf{} & \textbf{\rekall} \\ \cmidrule(l){2-5}
        \parking             & \textbf{98.6 $\pm$ 0.9}       & --                                  & --                    & 98.0            \\
        \bottomrule \\
        \vspace{-2em}
        \end{tabular}
        \caption{
        We validate the accuracy of \rekall\ queries against learned baselines.
        For classification tasks (top five rows) we
        train image classification networks (with and without temporal
        smoothing) and a temporal action recognition network as baselines and
        report F1 scores.
        We cast \parking\ as an object detection task and report average
        precision (AP) against a Faster R-CNN baseline.
        For all learned baselines, we report average scores and standard
        deviations over five random weight initializations.
        }
    \label{table:model_results}
\end{table*}

The goal of \rekall\ is to enable analysts to productively author
queries that meet the requirements of a range of video analysis tasks.
In this section, we evaluate \rekall\ in terms of the accuracy of the queries
compared to learned approaches and the usability of \rekall\ by domain experts
and novice users.
\begin{itemize}
\item
In Section~\ref{sec:evaluation_quality}, we evaluate our ability
to author high-accuracy \rekall\ queries
by comparing query output to that of learned baselines (deep models trained
on the same video data available to the programmer during query development).
In five out of six representative tasks, \rekall\ queries were on par with, and
in some cases significantly more accurate than, these learned baselines
(\AvgLift\ F1 points more accurate on average across all tasks, and up to
\MaxLift\ F1 points more accurate in one case).

\item
Since developing a program that detects an event of interest in a video
is more challenging than directly labeling instances of that event, 
we discuss the experiences of task domain experts authoring \rekall\ queries
for real-world video analysis applications 
(Section~\ref{sec:evaluation_user_study}).

\item
Finally, in a user study, we also evaluate how well novice \rekall\ users are
able to author \rekall\ queries for a new detection task in one hour.
Users with a sufficient functional programming background were able to author
queries that achieved average precision (AP) scores above \num{94} after
one-line tweaks by an expert user to account for errors in upstream object
detectors.
\end{itemize}

\subsection{Query Accuracy}
\label{sec:evaluation_quality}

We evaluate the accuracy of our \rekall\ queries by
comparing query outputs to baseline learning-based approaches on
six representative tasks for which high accuracy was required.
For each task, we collect human-annotated ground truth labels,
splitting labeled video into a \emph{development set} that was made
available to the \rekall\ programmer during query development,
and a held-out \emph{test set} used for evaluating the accuracy
of \rekall\ queries and trained models.
We train learning baselines using all human labels contained
in the development set.

\subsubsection{Classification Tasks}
\label{sec:classification_eval}

Five tasks (\interview,
\commercial, \conversation, \shotquery, and \shotscale)
can be viewed as classification tasks. 
For these tasks, we compare \rekall\
query accuracy (F1 score) against those of image
classification and action recognition networks trained on the
development set (results in Table~\ref{table:model_results}-top).  
These models classify whether a video frame or a short video segment
(for the action recognition baseline) contains the event of interest
(interview, commercial, etc.).
For the learned baselines, we report the average F1 score and standard
deviation over five random weight initializations.

\textbf{Setup}.
For the image classification baseline
(ResNet-50 Image Classification column in Table~\ref{table:model_results}),
we use a transfer learning approach\,\cite{pytorchtransferlearning} to
fine-tune a ResNet-50 image classifier\,\cite{ILSVRC15} (pre-trained on
ImageNet) for each task.
We also report the performance of this model after temporal ``smoothing":
taking the mode of model predictions over a window of seven frames
(ResNet-50 Image Classification + Smoothing).
Since \shotquery\ fundamentally requires information from multiple
frames to detect a shot change, we did not run the ResNet-50 baselines for this task.
For the action recognition baseline (Conv3D Action Recognition column),
we fine-tune a ResNet34-based 3D CNN (pre-trained on the Kinetics action
recognition dataset) for each task\,\cite{hara3dcnns}.
This network produces a single classification result given video
segments of $16$ frames.
We chose these methods to represent a breadth of ``reasonable-effort" learning
approaches to these classification problems, with a range of temporal
information available to the networks (the image classification baseline can
only see a single image, whereas the smoothing aggregates signal from a small
window, and the action recognition baseline is directly fed a larger window of
frames).

Details of the experimental setup for each classification task are given below.

\interview:
We limit the task to detecting interviews with Bernie Sanders
(and any host).
We annotated $\num{54}$ hours of TV news broadcasts,
split into a development set of $\num{25}$ hours, and a test set of
$\num{29}$ hours. Interviews with Sanders are rare;
the development set contains $\num{40}$ minutes of Sanders interviews,
and the test set contains $\num{52}$ minutes. 

\commercial:
We annotated commercials in $\num{46}$ hours of TV news
broadcasts, and partitioned this video into a development set of $\num{26}$ hours
($\num{7.5}$ hours of commercials), and a test set of $\num{20}$ hours
($\num{6.3}$ hours of commercials).

\conversation:
We annotated conversations in $45$ minutes of video 
selected from four films as a development set ($\num{29}$ minutes of conversations),
and annotated all conversations in a fifth film ($87$ minutes of footage,
$\num{53}$ minutes of conversations) as a test set.

\shotquery:
We annotated shot boundaries in $45$ minutes of video clips
randomly selected from $23$ movies, which we split in half
into a development set with $\num{348}$ shot transitions
and a test set with $\num{303}$ shot transitions.

\shotscale:
We annotated $898$ shots generated by the 
\shotquery\ query with scale annotations ($293$ long shots, $294$ 
medium shots, and $311$ close up shots).
We split these in half into a development set and a test set.

\textbf{Results}.
\rekall\ queries yielded a higher F1 score than the best learned model
in four of the five classification tasks (\AvgLift\ F1 points greater on
average across all five tasks).
The largest difference in accuracy was for
\shotscale, where the \rekall\ query was \MaxLift\ F1 points higher
than the best learned approach.

The performance of different learned approaches varied widely by
task; smoothing drastically improved the accuracy of the image classification
model for \interview, but much less so for \conversation\ and
\shotscale, and \textit{decreased} the accuracy for \commercial.
The action recognition baseline was the most accurate learning approach
for both \conversation\ and \shotscale, but was less accurate than the image
classification approaches for
\commercial\ and \num{77.8} F1 points lower than the \rekall\ 
query for \interview.

The learned baselines in Table~\ref{table:model_results} were chosen 
as reasonable-effort solutions that follow common practice.
It is likely that a machine learning expert, given sufficient time,
could achieve better results.
However, two of the tasks, \commercial\ and \shotquery, are well-studied and
have existing industrial or academic solutions.
We compared our \rekall\ commercial detector against that
of MythTV\,\cite{MythTV}, an open-source DVR system.
The MythTV detector achieved an F1 score of \num{81.5} on our test set,
\num{14.0} F1 points worse than the \rekall\ query.

For \shotquery, we compared against an open source implementation of the
DeepSBD shot detector\,\cite{hassanien2017deepsbd}, trained
on the large ClipShots dataset\,\cite{tang2018dsm}.
The DeepSBD shot detection model achieved an F1 score of
\num{91.4}, more accurate than our \rekall\ query.

However, by using the \rekall\ query's output 
on our entire \num{589} movie dataset
as a source of weak supervision\,\cite{Ratner16,Ratner18,Ratner19},
we are able to train a model that achieved an F1 score of
\num{91.3}, matching the performance of the DeepSBD model.
\emph{By using an imperfect \rekall\ query as a source of weak
supervision on a large, unlabeled video database,
we were able to train a model that matches the performance
of a state-of-the-art method using $636\times$ less ground truth data.}
For more details on this approach, see the \appendixortechreport.

\subsubsection{Object Detection Tasks}

In \parking\ we are interested both in detecting
\textit{when} there is an open parking spot, and
\textit{where} the open parking spot is in the frame.
Therefore, we cast it was an object detection problem; given an image of a
parking lot, detect all the empty parking spots.
We gathered two hours of parking lot footage,
annotated all the empty parking spots,
and split it into a development and test set.
We fine-tuned the Faster R-CNN model with ResNet-50
backbone\,\cite{ren2015faster,massa2018mrcnn} (pre-trained on MS-COCO\,\cite{COCO}) on the
development set.
The bottom row of Table~\ref{table:model_results} reports average precision
(AP) for \parking.
Both the learned baseline and the \rekall\ query are near-perfect, achieving
over \num{98.0} AP on this task.

\subsubsection{Query Performance}
Evaluation of deep neural network models often dominates the cost of video
analysis, so many recent video analysis systems focus on accelerating (or
avoiding) model execution\,\cite{Poms:2018:Scanner,kang2018blazeit,
kang2017noscope}.
In our application tasks, since the labels were pre-computed, \rekall\ queries
were able to run over the development sets quickly enough to enable iterative
query development (even though the \rekall\ implementation contains minimal
optimization).
Final queries for all tasks ran in less than thirty seconds on the development
sets.
The \rekall\ implementation stands to gain from further optimization, but even
naive implementations of the composition functions were sufficient to provide
interactive feedback to enable productive query development.
Since initial model evaluation is the primary computational cost, future
versions of \rekall\ might employ the various optimizations explored in
video analysis systems from the literature to reduce the initial cost of
pre-computing labels (or use query structure to guide cascades of model
evaluation).

\subsection{Usability}
\label{sec:evaluation_user_study}

\rekall\ programs have been authored by students in two university research
labs and at one company.
Our experiences suggest that many programmers can
successfully translate high-level video analysis tasks into queries.

For example, four of the representative tasks reported in
Section~\ref{sec:tasks} (\commercial, \conversation, \shotscale,
\filmidiom) were solved by domain experts
who had no previous experience using \rekall.
(\emph{Note: These users are now associated with the project and are
co-authors of this paper.}) 
These programming efforts ranged from an afternoon to
two days (time to learn how to use \rekall), and included overcoming common
query programming challenges such as dealing with 
noise in source annotations or misaligned transcript data. 

\begin{table}[t]
    \centering
    \small
    \begin{tabular}{@{}lrrrc@{}}
        \toprule
        \multicolumn{2}{c}{}  & \multicolumn{2}{c}{\textbf{AP Scores}}                     \\ \cmidrule(l){3-4} 
        \textbf{User ID} & \textbf{FP Experience} & \textbf{Original} & \textbf{Modified} \\ \midrule 
        1                & 5                      & 78.2              & 98.7              \\
        2                & 4                      & 75.0              & 98.7              \\
        3                & 3                      & 74.2              & 98.0              \\
        4                & 3                      & 66.5              & 95.5              \\
        5                & 3                      & 65.9              & 94.2              \\
        6                & 2                      & 66.5              & 95.5              \\
        7                & 1                      & 26.5              & 95.5              \\
        8                & 1                      & 0.0               & 0.0               \\
        \bottomrule \\
        \vspace{-2em}
        \end{tabular}
        \caption{
        Results of a user study with novice users on the
        \parking\ task.
        Participants were trained to use \rekall\ for one hour and then were
        given one hour to develop \rekall\ programs for the \parking\
        task.
        The Original column reports the average precision of their
        resulting programs.
        These scores were confounded by class confusion in the off-the-shelf
        object detector; scores after modification to account for this
        confusion (one LOC change for users 1-6, 3 LOC for user 7) are shown in
        the Modified column.
        Self-reported scores for familiarity with functional programming are
        shown in the FP Experience column.
        }
    \label{table:user_study_results}
\end{table}

These anecdotal experiences were promising, but the domain experts often
developed \rekall\ queries in close collaboration with \rekall\ developers.
To conduct a more quantitative assessment of the usability of \rekall, we ran a
user study to evaluate how well novice users of \rekall\ were able to author
queries to detect new events given one hour of query development time.

We recruited \num{eight} students
with backgrounds ranging from medical
imaging to machine learning and computer architecture.
Participants received a one-hour \rekall\ training session, and then were
given one hour to write a query for the \parking\ task, along with a high-level
English-language description of an approach similar to the one described in
Section~\ref{sec:tasks}.
Six of the eight students were able to successfully create a label set
representing parking spaces, and use the \code{minus} operation to subtract car
detections.
Although their queries were algorithmically similar to our results, none of the
students was able to achieve comparable accuracy to our solution, since they
did not account for cross-class confusion (cars being mis-classified as
trucks).
After modifying participant queries to account for these failures, user queries
achieved average precision scores above $\num{94}$.

The average precision scores of participant queries are shown in
Table~\ref{table:user_study_results}; the results from the original queries are
shown in the Original column.

Users 1-6 were able to successfully subtract car detections from a label set
representing parking spaces.
Of those six, three (users 1-3) had enough time remaining to further iterate
by using the \code{coalesce} operation to smooth over noise in the output
of the object detector.
This increased the AP scores of their
algorithms by an average of \num{11.5} points.
User~7 developed a query computing a \code{minus} operation between proposed
parking spots and car detections, but did not correctly construct the parking
spots event set; this resulted in an AP score of \num{26.5}.
User~8 failed to use \rekall\ to construct the initial parking spot
label set, which was later determined to be due to a lack of familiarity
with functional programming interfaces.

After the study, we manually modified each student’s solution to understand how
far they were from optimal.
The results from the modified queries are shown in the Modified
column.
Our modifications relaxed the \code{minus} operation to subtract out more
objects, and not just the detected cars.
For users 1-7, this was a single line change (equivalent of line
$8$ in Figure~\ref{lst:parking}).
For user 7, we additionally fixed the incorrect parking spot construction
(equivalent of line $5$ in Figure~\ref{lst:parking}).
After our changes, the queries written by users 1-7 all achieved AP scores
greater than $94$.
The modified versions of the queries written by
users 1 and 2 were more accurate than our algorithm for \parking.
This suggests that the participants were close, and with a better
understanding of computer vision failure modes and debugging tools they could
likely have achieved near-perfect accuracy.

Success using \rekall\ was strongly correlated with participants'
self-rated familiarity with functional programming.
Before the tutorial, we asked users to rate their familiarity with
functional programming on a scale of $1-5$, with $1$ being ``Not at all
familiar" and $5$ being ``Most familiar."
The self-reported scores are shown in column FP Familiarity in
Table~\ref{table:user_study_results}.

\section{Related Work}
\label{sec:related}

\textbf{Multimedia Database Systems}.
The idea of associating database records with video intervals and composing
queries for video search goes back to multimedia database systems from the
90's and early 2000's.
Systems such as OVID, MMVIS, AVIS, CVQL, and
BilVideo aimed to provide an interface to query for complex events in
video\,\cite{Donderler:2004:bilvideo,
Adali:1996:avis,Koprulu:2002:avis-st,hibino:1995:mmvis-tvql,oomoto:1993:ovid,
Kuo:1996:CVQL,Li1997ModelingOM}.
The query languages for these systems supported spatiotemporal joins based on
Allen interval operations\,\cite{Allen:1983:interval} and included operations
to express repetitions of patterns.

However, queries written in these systems lacked the starting point of a large
set of useful video annotations that modern machine learning technologies can
provide.
The modern machine learning landscape now makes it possible for heuristic
composition to quickly define new events.
We view our efforts as adapting early ideas about spatiotemporal
composition to a modern context, with modern primitive annotations, large video
datasets, and new, real-world video analysis tasks.

\textbf{Domain-Specific Video Retrieval Systems}.
Our work
is related to work on domain-specific video retrieval systems, such as
SceneSkim or RoughCut in the film domain \cite{ronfard2003,Lehane:2007,
mohamad2011,Pavel:2015,Leake:2017:roughcut,Wu:2018:filmpatterns,
Wu:2016:constrainedPatterns,Wu:2017:filmstyle} or Chalkboarding in the sports
domain \cite{sha2016chalkboarding,sha2017fine,power2017not,
power2018mythbusting}.
These systems take advantage of domain-specific structure to support efficient
video retrieval.

These approaches are largely complementary to programmatic composition; many of
them break down video streams into domain-specific events, and allow users to
query over sequences of those events.
\rekall\ could be used to express or model these domain-specific events, and
the broader
patterns could be broken down into a series of composition
operations.

\textbf{Complex Event Processing and Diverse Analytics Systems}.
\rekall's composition operations takes inspiration from complex event
processing and temporal stream analysis systems,
such as Apache Flink, SiddiQL, and Microsoft
Trill\,\cite{ApacheFlink, SiddhiQL, chandramouli2015trill}.
Adapting operators from the complex event processing systems to the video
domain required support for a few language features that are not universal
among complex event processing systems.
In particular, we found that a continuous interval representation instead of a
point representation was necessary to model data from different modalities,
sampled at different resolutions.
Complex event processing systems such as Apache Flink and SiddiQL are based on
point events only, so they lack many of the interval-based operations that we
have found necessary for video event specification.
Trill provides a rich set of operations for analytics on data streams,
including interval-based operations and anti-semi joins, so its expressivity is
similar to \rekall.
The one operation that does not appear directly is an equivalent of
\code{coalesce}.
It would be interesting to consider how systems like Trill could be adapted for
future video event specification tasks. 

\textbf{Few-Shot and Zero-Shot Learning}.
Our approach
is also related to recent efforts in
few-shot and zero-shot learning \cite{snell2017prototypical,
lampert2013attribute,socher2013zero,jain2015objects2action,gan2016you,
yang2018one}.
In fact, one way to view programmatic composition is as a mechanism for
few-shot or zero-shot event detection in video, where the query programmer uses
a small number of examples to guide query development.
The mechanics of the approach to writing a \rekall\ query are different from
most approaches to few-shot or zero-shot learning, but some of the principles
remain the same; for instance, programmatic composition relies on information from
pre-trained networks in order to compose them into complex events.

\rekall\ queries could also be used in conjunction with few-shot learning
approaches.
In many of our scenarios, \rekall\ queries are used for video event
retrieval when there are no examples, only an idea in a developer's head.
In these cases, initial \rekall\ queries with human-in-the-loop curation could
be used to source an initial small amount of labeled data to bootstrap few-shot
learning approaches.

\textbf{Weak Supervision Systems}.
We have shown that it is possible in a number of situations to use \rekall\ to
write queries that accurately detect new events of interest, but it may not
always be possible to programmatically specify new events accurately enough for
all downstream applications.
Weak supervision systems such as Snorkel\,\cite{Ratner16,Ratner18,
Ratner19} and Coral\,\cite{varma2017coral} use statistical techniques to
build accurate models when accurate heuristics may be difficult to author.
One use of \rekall\ is as a mechanism for writing richer heuristics for these
weak supervision systems when they are operating over video;
in fact, we utilize weak supervision techniques in the
\shotquery\ task to weakly supervise a model that is more accurate than a
model trained on $636\times$ more ground-truth data than we had access to.
More work may be required to adapt weak supervision techniques to the video
domain\,\cite{khattar2019gait};
in the future, we plan on exploring how to more deeply integrate \rekall\
queries into weak supervision systems.

\textbf{Video Retrieval Through Natural Language Interfaces}.
Some recent works at the intersection of computer vision and natural language
processing have centered around the task of action localization through natural
language interfaces \cite{hendricks17iccv,gao2017tall}.
These approaches aim to solve a similar problem to the video event
specification problem that we consider, but the setup is slightly
different; they are limited to the set of natural language descriptions found
in the training/test distributions.
This is fine if the events of interest are ``in distribution," but becomes
problematic for more complex events.
For example, the query ``Jake Tapper interviews Bernie Sanders" would fail
unless the network were trained on a broad enough set of queries to understand
who Jake Tapper and Bernie Sanders were, and to create an embedding for the
concept of an interview.
The programmatic approach to event specification allows analysts to encode
complex concepts directly by using domain knowledge to compose simpler concepts
together.

\section{Discussion}
\label{sec:discussion}

\rekall\ is intended to give analysts a new tool 
for quickly specifying video events of interest
using heuristic composition.
Of course,
the notion of authoring code in a domain-specific query language
is not new, but adopting this approach for video analysis
contrasts with current trends in modern machine learning,
which pursue advances in video event detection through
end-to-end learning from raw data (e.g. pixels, audio,
text)\,\cite{hara3dcnns}.
Constructing queries through procedural
composition lets users go from an idea to a set of
video event detection results rapidly, does not incur the costs of
large-scale human annotation and model training,
and allows a user to express heuristic domain knowledge
(via programming), modularly build on existing labels,
and more intuitively debug failure modes.  

However, compositional video event specification has many known limits.
\rekall\ queries still involve manual
parameter tuning to correctly set overlap or
distance thresholds for a dataset.
Higher-level composition is difficult when lower-level labels do not exist or
fail in a particular context.
(Our film studies efforts failed to build a reliable kissing scene detector 
because off-the-shelf face and human pose detectors failed due to occlusions
present during an embrace.)
In future work we plan to pursue \rekall\ extensions
that model richer description of human behaviors or fine-grained movements,
but there will always be video events that are less amenable to
compact compositional descriptions and better
addressed by learned approaches.

Nevertheless, we believe productive systems for compositional video event
specification stand to play an important role in the development
of traditional machine learning pipelines
by helping engineers write programs that surface a more diverse set of
training examples for better generalization,
enabling search for anomalous model outputs (feeding active learning loops),
or as a source of weak supervision to bootstrap model training. 
We hope that our experiences encourage the community to explore 
techniques that allow video analysis efforts to more effectively
utilize human domain expertise and more seamlessly provide solutions 
that move along a spectrum between traditional query programs and learned models.

\iftechreport
\section*{Acknowledgments}
\noindent
\small{We thank Jared Dunnmon, Sarah Hooper, Bill Mark, Avner May, and Paroma Varma
for their valuable feedback.
We gratefully acknowledge the support of DARPA under Nos. FA87501720095 (D3M),
FA86501827865 (SDH), and FA86501827882 (ASED), NIH under No. U54EB020405
(Mobilize), NSF under Nos. CCF1763315 (Beyond Sparsity), CCF1563078 (Volume
to Velocity),
1937301 (RTML),
III-1908727 (A Query System for Rapid Audiovisual Analysis of Large-Scale Video
Collections),
and III-1714647 (Extracting Data and Structure from Charts and Graphs for
Analysis Reuse and Indexing),
ONR under No. N000141712266 (Unifying Weak Supervision), the
Moore Foundation, NXP, Xilinx, LETI-CEA, Intel, Microsoft, NEC, Toshiba, TSMC,
ARM, Hitachi, BASF, Accenture, Ericsson, Qualcomm, Analog Devices, the Okawa
Foundation, and American Family Insurance, Google Cloud, Swiss Re,
Brown Institute for Media Innovation,
Department of Defense (DoD) through the National Defense Science and
Engineering Graduate Fellowship (NDSEG) Program, 
and members of the Stanford DAWN
project: Teradata, Facebook, Google, Ant Financial, NEC, SAP, VMWare, and
Infosys.
The U.S. Government is authorized to reproduce and distribute reprints for
Governmental purposes notwithstanding any copyright notation thereon.
Any opinions, findings, and conclusions or recommendations expressed in this
material are those of the authors and do not necessarily reflect the views,
policies, or endorsements, either expressed or implied, of DARPA, NIH, ONR, or
the U.S. Government.}

\fi
\bibliographystyle{abbrv}
\bibliography{rekall_arxiv} 
\iftechreport
\clearpage
\section{Appendix}

\subsection{Supplemental Videos}
We provide eight supplementary videos on our paper website
(\url{http://www.danfu.org/projects/rekall-tech-report/}):
\begin{itemize}
\item
Film trailer for \textit{Star Wars: Episode III - Revenge of the Sith}.
We used \rekall\ queries to mine for clips matching eleven different film
idioms, based on a viral video ``How To Make a Blockbuster Movie Trailer."
We manually selected a few and edited them together into a trailer according to
a template.

\item
Video montage of TV News interviews.
We used our interview query to find interview clips with the 2016 presidential
candidates.
We randomly sampled $64$ clips from the results and edited them into a video
montage.

\item
Supercut of reaction shots from \textit{Apollo 13}.
We wrote a query to find reaction shots -- shots where the character onscreen
is shown reacting to another (offscreen) character's dialogue.
We ran the query on \textit{Apollo 13} and put together a supercut of all the
results.

\item
Supercut of action sequences.
We wrote a query to find action sequences, and put together a supercut of a
sample of the results.

\item
Supercut of ``May The Force Be With You."
We wrote a query to find all the times when the phrase ``May The Force Be With
You" is said in the \textit{Star Wars} films, and put together a supercut of
all the results.

\item
Supercut of Hermione between Ron and Harry.
We wrote a query to find the spatial pattern of Hermione Granger between Ron
and Harry in the \textit{Harry Potter} series, and put together a supercut of
all the results.

\item
Film idiom: Intensification.
We wrote a query to find instances of the ``intensification" film idiom -- where
the shot scale gets monotonically larger on one or both characters throughout
the course of a conversation.
We selected one instance from the result set to show (from the movie
\textit{Spellbound}).

\item
Film idiom: Star Wide.
We wrote a query to find instances of the ``start wide" film editing idiom --
where the first shot in a conversation shows all the characters in the
conversation.
We selected one instance from the result set to show (from the movie
\textit{Interstellar}).
\end{itemize}

\subsection{Additional Task Details}
In this section, we provide additional code listings and details for the
\commercial, \shotquery, \conversation, and \filmidiom\ tasks
discussed in Section~\ref{sec:tasks}.

\subsubsection{Commercial Detection}
\begin{figure}[ht!]
\begin{lstlisting}
# Commercial Query
transcript_words = rekall.ingest(transcript, 1D)
histograms = rekall.ingest(database.table("hists"), 1D)
entire_video = rekall.ingest(database.table("video"), 3D)

transcripts_with_arrows = transcript_words
  .filter($\lam$ word: '>>' in word)

black_frame_segs = histograms
  .filter($\lam$ i: i.histogram.avg() < 0.01)
  .coalesce(predicate = time_gap < 0.1s, merge = time_span)
  .filter($\lam$ i: i["t2"] - i["t1"] > 0.5s)

candidate_segs = entire_video.minus(black_frame_seqs)

non_commercial_segs = candidate_segs
  .filter_against(
    transcripts_with_arrows,
    predicate = time_overlaps)

commercial_segs = entire_video
  .minus(non_commercial_segs.union(black_frame_segs))

commercials = commercial_segs
  .coalesce(predicate = time_overlaps, merge = time_span)
  .filter($\lam$ i: i["t2"] - i["t1"] > 10s)

lower_case_word_segs = transcript_words
  .filter($\lam$ word: word.is_lowercase())
  .coalesce(predicate = time_gap < 5s, merge = time_span)

missing_transcript = entire_video
  .minus(transcript_words)
  .filter($\lam$ i: 30 < i["t2"] - i["t1"] < 270)

commercials = commercials
  .union(lower_case_word_segs)
  .union(missing_transcript)
  .coalesce(predicate = time_gap < 45s, merge = time_span)
  .filter($\lam$ comm: comm["t2"] - comm["t1"] < 300s)
\end{lstlisting}
\caption{
A Rekall query to retrieve commercials in a video collection of TV News
broadcasts.
}
\label{lst:commercials}
\end{figure}

A code listing for the \commercial\ query is shown in
Figure~\ref{lst:commercials}.
In our dataset, commercials are delineated by sequences of black frames.
They often have mixed-case or missing transcripts, while the transcripts of
non-commercial segments typically contain upper case text and distinct speaker
markers (\code{>>}).
The commercial detection query takes advantage of these dataset-specific
heuristics to partition the video by sequences of black frames
(lines~\num{9-14}), filter out segments that intersect with speaker markers
(lines~\num{16-22}), and add in segments where the transcripts are mixed-case
or missing (lines~\num{28-40}).
The \code{filter_against} join, used in line~\num{16}, performs a left outer
join and keeps any label from the left set that passes the predicate with any
label from the right set.
This query also makes heavy use of the \code{coalesce} function to connect
sequences of labels that are temporally disconnected.

\subsubsection{Shot Transition Detection}
\begin{figure}[ht!]
\begin{lstlisting}
# Shot Query
histograms = rekall.ingest(database.table("hists"), 1D)

hist_pairs = histograms
  .join(
    histograms,
    predicate = after(1 frame),
    merge = time_span)

hist_diffs = hist_pairs
  .map($\lam$ pair: pair.first
      with payload = diff(pair.first, pair.second))

windows = hist_diffs
  .join(
    hist_diffs,
    predicate = before or after 500 frames,
    merge = $\lam$ a, b: a with payload = [b])
  .coalesce(
    predicate = time_equal,
    merge = $\lam$ a, b:
      a with payload = a.payload + b.payload)
  .map($\lam$ window:
    window with payload = {
      diff: window.interval.diff,
      avg: mean(window.payload),
      stddev: stddev(window.payload)
    })

transitions = windows
  .filter($\lam$ i: i.diff > 2.5 * i.stddev + i.avg)
  .coalesce(
    predicate = time_gap < 10 frames,
    merge = time_span)
  .map($\lam$ t: t with t["t2"] = t["t1"] + 1 frame)

faces = rekall.ingest(database.table("faces"), 3D)

faces_by_frame = faces.group_by($\lam$ face: face["t1"])

transitions_starting_faces = transitions
  .join(
    faces,
    predicate = start_time_equal,
    merge = $\lam$ transition, faces:
      transitions with payload = faces)

transitions_ending_faces = transitions
  .join(
    faces,
    predicate = end_time_equal,
    merge = $\lam$ transition, faces:
      transitions with payload = faces)

bad_transitions = transitions_starting_faces
  .join(
    transitions_ending_faces,
    predicate = time_equal and faces_match,
    merge = $\lam$ t1, t2: t1)

transitions = transitions.minus(bad_transitions)
\end{lstlisting}
\caption{
A Rekall query to detect shot transitions in film.
}
\label{lst:shotquery}
\end{figure}

A code listing for the \shotquery\ query is shown in
Figure~\ref{lst:shotquery}.
This query uses color histograms of each frame to find sudden changes in the
frame content.
It computes the differences between histograms of neighboring frames and finds
sudden spikes in the differences by computing the average and standard
deviation of the change over a large window of frames.
The query further reduces false positives by using the positions of face
detections at the beginning and end of each shot transition; in short, if the
query finds the same number of faces in the same positions at the beginning and
end of a shot transition, it removes the transition as a false positive.

Lines~\num{4-12} compute differences between the histograms in neighboring
pairs of frames.
Lines~\num{14-28} construct a sliding window over the entire film, and compute
the average and standard deviation of between-frame changes for each window.
Lines~\num{30-35} isolate frame pairs where the change in frame histogram
is at least \num{2.5} standard deviations greater than the average change in
the entire window, and remove transitions that are within \num{10} frames of
any other transitions.
Lines~\num{41-53} associate faces with the beginning and end of each
transition, and lines~\num{55-59} use the custom \code{faces_match} predicate
to detect transitions where the faces at the beginning and end of each
transition are in the same place (just comparing the count and position of the
nested face detections).
Finally, line~\num{63} removes the bad transitions from the pool.

For sake of exposition, we have presented this query as detecting shot
transitions in a film (i.e., the frame(s) at which a shot changes).
This is how we evaluate the accuracy of a shot detector (how many of the
detected transitions represent true transitions), but we store shot information
in terms of continuous shots (i.e., we store a single label for each contiguous
shot, instead of a label for each shot transition).
The conversion is simple; a \code{minus} operation can partition a film into
contiguous shots to be stored for future use (such as for the \shotscale\ or
\conversation\ queries).

\subsubsection{Conversation Detection}
\begin{figure}[ht!]
\begin{lstlisting}
# Conversation Query
faces = rekall.ingest(database.table("faces"), 3D)
shots = rekall.ingest(database.table("shots"), 1D)

faces_by_frame = faces.group_by($\lam$ face: face["t1"])
shots_with_face_clusters = shots
  .join(
    faces_by_frame,
    predicate = time_overlaps,
    merge = span)
  .group_by($\lam$ shot: (shot["t1"], shot["t2"]))
  .map($\lam$ shot:
    shot with shot.payload = cluster_faces(shot.faces))

shot_pairs = shots_with_face_clusters
  .join(
    shots_with_face_clusters,
    predicate = $\lam$ s1, s2: s1["t2"] == s1["t1"],
    merge = span)

shot_pair_sequences = shot_pairs
  .coalesce(
    predicate = time_overlaps and
      $\lam$ shot_pair1, shot_pair2:
        True if $\exists$ tup1 $\in$ shot_pair1.id_pairs,
                  tup2 $\in$ shot_pair2.id_pairs |
            (dist(tup1[0], tup2[0]) < THRESHOLD and
             dist(tup1[0], tup2[0]) < THRESHOLD) or
            (dist(tup1[0], tup2[1]) < THRESHOLD and
             dist(tup1[1], tup2[0]) < THRESHOLD),
    merge = span
  )

adjacent_sequences = shot_pair_sequences
  .join(
    shot_pair_sequences,
    predicate = after one frame and
      $\lam$ shot_pairs1, shot_pairs2:
        true if $\exists$ tup1 $\in$ shot_pairs1.id_pairs,
                  tup2 $\in$ shot_pairs2.id_pairs |
            (dist(tup1[0], tup2[0]) < THRESHOLD and
             dist(tup1[0], tup2[0]) < THRESHOLD) or
            (dist(tup1[0], tup2[1]) < THRESHOLD and
             dist(tup1[1], tup2[0]) < THRESHOLD),
    merge = span)

conversations = shot_pair_sequences
  .union(adjacent_sequences)
  .filter(num_shots >= 3)
  .coalesce(predicate = time_overlaps, merge = span)
\end{lstlisting}
\caption{
A Rekall query to detect conversations in film.
}
\label{lst:conversations}
\end{figure}

A code listing for the \conversation\ query is shown in
Figure~\ref{lst:conversations}.
This query makes heavy use of a simple heuristic for detecting conversations --
shots of the same people appearing multiple times in a contiguous film segment.
In our dataset, caption data was difficult to obtain (manual searching and
scraping repository websites), and was often unreliable (missing or badly
mis-aligned with the video files).
As a result, we decided to write our query entirely from visual content.
Since films are highly stylized, we couldn't rely on off-the-shelf identity
recognition (which may have failed on minor characters).
Instead, we used distance in face embedding space as a simple proxy for whether
two detected faces belonged to the same character or not.

The query starts by loading up face detections and embeddings from
off-the-shelf face detection and embedding networks\,\cite{schroff2015facenet,
zhang2016joint}, and contiguous shots output by the \shotquery\ query
(lines~\num{2-3}).
It then clusters the faces in a single query by the face embeddings, using the
custom \code{custom_faces} map function, which just uses K-means clustering
with K equal to the maximum number of faces in a single frame in the shot
(lines~\num{5-13}).
Next, the query constructs pairs of neighboring shots (lines~\num{15-19}), and
uses the \code{coalesce} function to construct sequences of shot pairs where
the same people appear throughout the sequence, expressed by a custom predicate
function (lines~\num{21-32}).
Finally, the query joins together neighboring sequences, eliminates any
sequences that are fewer than three shots long -- i.e., the ones that are still
shot pairs after the sequence construction (lines~\num{47-50}).

Conversation detection algorithm in Figure~\ref{lst:conversations}.
English-language description of the steps here.

\subsubsection{Film Idiom Mining}
In this section, we present a full list of film idioms we mined for (including
for the film trailer and supplementary videos), and give a brief description of
the \rekall\ queries used to mine for these idioms.
Many of these queries operate solely on face or object detections and
transcript data, although a few also utilize identity information for main
characters or the main villain.
Before starting the mining process for the movie, we cluster the faces in a
movie based on the face embeddings to quickly label all the main characters and
the main villains in the movie --  a process which took less than ten minutes
per movie on average.

The film trailer template contained a total of six different film idioms that
we mined for:
\begin{itemize}
\item
Establishing shot -- query searched for shots with no people or objects
\item
Ponderous statement or question -- query searched the transcript for phrases
like ``what if?"
\item
Action sequences -- query searched for sequences of multiple short shots in a
row
\item
Dark tidings -- query searched for sequences where the main villain was shown
on screen along with some text in the transcript
\item
Hero shots -- query searched for shots where a main character was displayed
alone in a medium shot or close up, and the brightness of the frame was high
(manually-set threshold)
\item 
Shots of main characters looking hopeful -- query was the same as the hero
shots query, except allowing multiple characters in the frame
\end{itemize}
The trailer template contained a few other items as well, but we found that
we could re-use existing queries for these.
For example, the trailer template contains ``statement of causality or finality
from the main villain" as well as ``dark tidings from the main villain," but we
were able to use the same dark tidings query for both.

\subsection{Weak Supervision}
In Section~\ref{sec:evaluation_quality}, we briefly discussed using \rekall\
queries as a source of weak supervision to train a model on a large collection
of unlabeled data for the \shotquery\ task.
In this section, we discuss this approach in more detail.

The \rekall\ algorithm for the \shotquery\ task was developed on a relatively
small selection of data (half of 45 minutes of ground truth data that we
collected), since collecting ground truth shot transition labels is expensive
(labelers need to inspect every frame in a sequence to find the frame where a
shot transition takes place).
However, we had access to a much larger collection of unlabeled data -- $589$
feature-length films.
We wanted to use the signal in the unlabeled data to improve over the
performance of our \rekall\ query.

We turned to open-source \textit{weak supervision} techniques in the academic
literature such as Snorkel\,\cite{Ratner16, Ratner18, Ratner19}.
These techniques provide a mechanism to label a large collection of unlabelled
data by using statistical methods to estimate the accuracies of different
\textit{labeling functions}.
A labeling function is a function -- any piece of code -- that can label a
data point.
Weak supervision systems like Snorkel use the observed agreements and
disagreements between labeling functions \textit{on unlabelled data} to
estimate the underlying accuracies of the labeling functions.
The accuracies and labeling function outputs can then be used to output
probabilistic labels over the unlabelled data, which can be used to train an
end model using standard training techniques.
To summarize, these weak supervision systems use multiple noisy user-defined
labeling functions to generate probabilistic training labels for a collection
of unlabelled data.

In order to apply these techniques to the \shotquery\ task, we needed to turn
our single query into multiple labeling function.
The \shotquery\ query can be
viewed as a labeling function, but there is no way for weak supervision systems
to observe agreements and disagreements between labeling functions if there is
only a single labeling function.

To overcome this challenge, we broke the \shotquery\ query into five different
labeling functions -- three that looked at changes in histograms between
frames, and two that looked at the number and position of faces between frames.
The three labeling functions that looked at changes in histograms between
frames were all equivalent to lines~\num{2-35} in Figure~\ref{lst:shotquery},
but used different histograms as input -- one used RGB histograms, one used
HSV histograms, and one used optical flow histograms.
The two face detection labeling functions, meanwhile, distilled the domain
knowledge from lines~\num{41-59} in Figure~\ref{lst:shotquery} and voted that
there was no shot transition if they detected the same number of faces, or
faces in the same position, respectively, across a series of frames.
These labeling functions were all written in \rekall\ and generated noisy
labels for the entire film dataset.

Once we generated these noisy labels, we used the weak supervision techniques
in the Snorkel system\,\cite{Ratner18} to learn the accuracies of the five
labeling functions from their agreements and disagreements and generate
probabilistic labels over the full unlabelled dataset.
We then used these labels to train a shot detector, which achieved an F1 score
of \num{91.3} on the held-out test set -- matching the performance of another
shot detector trained on a collection of gold ground-truth data $636\times$
larger than our ground-truth dataset.

\fi

\end{document}